\documentclass[conference]{IEEEtran}
\usepackage[subpreambles=true]{standalone}
\usepackage{3-way-intersect}
\usepackage{import}

\graphicspath{{figures/}}

\begin{document}
	\title{Learning Driver Behaviors Using A Gaussian Process Augmented 
	State-Space Model}
	\author{\IEEEauthorblockN{Anton Kullberg\\\small{Linköping 
				University}\\anton.kullberg@liu.se}
		\and
		\IEEEauthorblockN{Isaac Skog\\
			\small{Linköping University}\\isaac.skog@liu.se}
		\and
		\IEEEauthorblockN{Gustaf Hendeby\\
			\small{Linköping University}\\gustaf.hendeby@liu.se}}

\maketitle
\begin{abstract}
	An inference method for Gaussian process augmented state-space models
	are presented. This class of grey-box models enables domain knowledge
	to be incorporated in the inference process to guarantee a minimum of
	performance, still they are flexible enough to permit learning of
	partially unknown model dynamics and inputs. To facilitate online
	(recursive) inference of the model a sparse approximation of the Gaussian
	process based upon inducing points is presented. To illustrate the
	application of the model and the inference method, an example where it
	is used to track the position and learn the behavior of a set of cars
	passing through an intersection, is presented. Compared to the case
	when only the state-space model is used, the use of the augmented
	state-space model gives both a reduced estimation error and bias. 
\end{abstract}

\section{Introduction}

State estimation is a fundamental problem in many areas, such as robotics, 
target tracking, economics, etc. Typical techniques for state estimation are 
Bayesian filters such as the Kalman Filter (\abbrKF), Extended Kalman Filter 
(\abbrEKF) or the Particle Filter (\abbrPF) 
\cite{Sarkka2010:BayesianFiltering}. Common to these techniques is that 
they require a model for the state dynamics as well as an observation model for 
the considered system. A general model for this is
\begin{subequations}
	\label{eq:introduction:generalmodel}
	\begin{align}
		\state_k &= \vec{f}_k(\state_{k-1},\vec{u}_{k},\vec{w}_{k})\\
		\obs_k &= \vec{h}_k(\state_k,\vec{e}_k),
	\end{align}
\end{subequations}
where $\vec{f}_k$ and $\vec{h}_k$ are the state dynamics and observation model, 
respectively, and can be both time-varying and nonlinear. Further, $\state_k$ 
is the state vector at time $k$, $\vec{u}_k$ is some input, and $\vec{w}_k$ and 
$\vec{e}_k$ are mutually independent Gaussian white noise with covariance 
matrices $Q$ and $R$, respectively. Commonly, $\vec{f}_k$ and $\vec{h}_k$ are 
chosen based upon insight about the underlying physical process and the sensors 
used to observe it \cite{LjungGlad:ModelingIdentification}. \par
If information about the underlying process is missing, a model can be inferred 
from data using system identification methods 
\cite{LjungGlad:ModelingIdentification}. This 
can for instance be done using a Gaussian Process State-Space Model 
(\abbrGPSSM), where both the state dynamics and observation model are viewed as 
separate Gaussian Processes ({\abbrGP}s) and learned jointly 
\cite{Svensson2016, Svensson2017, Turner2010:SSInferencewithGP, Frigola2013, 
Ko2009}. These methods may include prior physical knowledge, but only through 
the prior mean and is thus not so flexible. Moreover, they are computationally 
demanding and not well suited for online application.
\subsection{Augmented State-Space Models}
In many situations, knowledge about parts of the underlying process might 
exist, or a simplified model of the process exists. One way to incorporate 
this knowledge into the state estimation problem is to separate the system into 
two parts, one known (here assumed linear) and one unknown part, where the 
unknown part is modeled as a \abbrGP \cite{Veiback2019:EKFGP}. That is,
\begin{subequations}
	\label{eq:introduction:basemodel}
	\begin{align}
	{}&\vec{f}_k(\state_{k-1},\vec{u}_k,\vec{w}_k) \simeq 
	F_k\state_{k-1}+G_k\vec{u}_k+G_k\vec{w}_k\\
	{}&\vec{u}_k = \begin{bmatrix}
		u^1(\vec{z}_k) & u^2(\vec{z}_k) & \cdots & u^J(\vec{z}_k)
	\end{bmatrix}^T\\
	{}&u^j(\vec{z}) \sim \mathcal{GP}(0,k(\vec{z},\vec{z}')).
	\end{align}
\end{subequations}
Here, $\vec{u}_k$ is an unknown input at time $k$, which is to be inferred from 
data. It depends on some variable $\vec{z}_k$, which may change over time. 
For example, $\vec{z}_k$ could be a function of the state $\vec{x}_k$. 
Furthermore, the input $\vec{u}_k$ is modeled as a \abbrGP but could also, 
e.g., be a basis function expansion or neural network \cite{Svensson2015, 
Sjoberg1994}. Henceforth, the model \eqref{eq:introduction:basemodel} will be 
referred to as a Gaussian Process Augmented State-Space Model (\abbrGPASSM). 
The benefits of this class of grey-box models are that they provide:
\begin{enumerate}
	\item \textbf{Performance guarantees} --- the model will perform at least 
	as well as the simplified state dynamics model.
	\item \label{bp:introduction:interpretability}\textbf{Interpretability of 
	$\vec{u}$} --- in many cases, the learned inputs have a meaningful, 
	physical interpretation.
	\item \textbf{Improvement over time} --- as more data is available, the 
	model accuracy improves and so also the state estimation accuracy.
	\item \textbf{Information sharing} --- if many systems behave similarly, 
	the input $\vec{u}$ may be shared between systems and learned jointly.
\end{enumerate}
This class of models could for instance be exploited for tracking ships, or 
icebergs, where the learned input could be interpreted as the ocean currents 
\cite{Veiback2019:EKFGP}. It could also be used for estimating the structural 
integrity of buildings which is affected by unknown outer forces, which could 
be learned \cite{Nayek2019}. A final example is predicting the number of 
remaining charging cycles of batteries. In Lithium-Ion batteries it is well 
understood that the usage history and environment affect the degradation 
process \cite{Ecker2014}. These effects can potentially be learned from other 
batteries with similar history to accurately predict degradation and schedule 
maintenance accordingly.
\subsection{Scenario Description}
\begin{figure}[ht]
	\centering
	\includegraphics[width=\columnwidth]{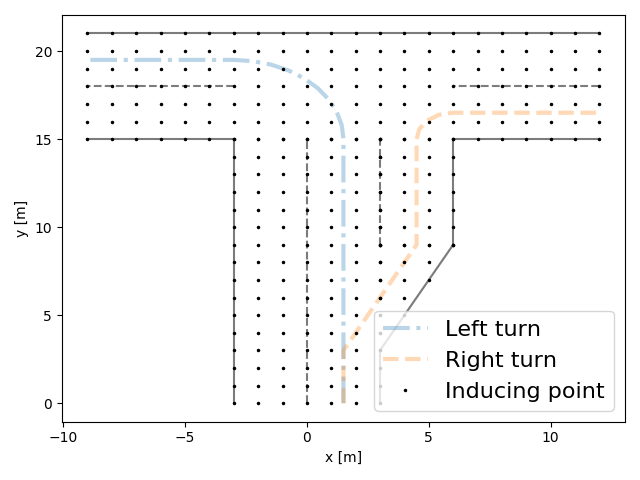}
	\caption{Considered intersection scenario. A vehicle approaches from the 
		middle road and may, with equal probability, travel along either of the 
		two paths, i.e., either turn left or right. The inducing points are 
		fixed on a grid covering the entire intersection.}
	\label{fig:description:map}
\end{figure}
Before an estimator for the \abbrGPASSM is derived, a usage scenario is 
presented to give a foundation for the forthcoming discussion. The 
considered scenario is visualized in Fig \ref{fig:description:map}. A vehicle 
approaches the intersection from the middle road and travels through the 
intersection according to one of the two possible paths. The problem is to 
estimate the state of the vehicle given by 
\begin{equation}
\label{eq:scenario:states}
\state_k=\begin{bmatrix}
p_x & p_y & v_x & v_y
\end{bmatrix}^T,
\end{equation}
where $p_i$ and $v_i$ is the position and velocity along the 
$i$:th-axis, respectively. In a typical vehicle tracking application, the 
information provided by its trajectory is forgotten and the trajectory of each 
new vehicle passing through the intersection is assumed independent from its 
predecessors. In scenarios such as that illustrated by Fig 
\ref{fig:description:map}, this assumption is naive since each vehicle in the 
intersection are bound to follow traffic regulations, the road network, etc. 
Hence, there will be a high correlation between the dynamics of the vehicles 
that pass through the intersection. This dynamic can be modeled as an 
unknown input to the model of the vehicles' motion and can be learned by 
sharing information between vehicles. To that end, the motion dynamics of the 
vehicles will be modeled by a \abbrGPASSM where the known model dynamics is 
given by a Constant Velocity (\abbrCV) model \cite{Li2003}, and the unknown 
driver input, i.e., the accelerations, are modeled by a \abbrGP.


\section{Modeling}\label{sec:modeling}

In its basic form, the \abbrGPASSM presented in 
\eqref{eq:introduction:basemodel} cannot be used for online learning and state 
estimation as the inference of the \abbrGP increases cubically with the number 
of observations \cite{RasmussenWilliams2006:GPforML}. Therefore, we will next 
introduce a sparse approximation for 
the Gaussian process and modify the model \abbrGPASSM accordingly. For 
simplicity, but without loss of generality, the observation function will be 
assumed linear.      

\subsection{Approximating the input}
There has been a lot of research in approximating 
\abbrGP regression, both in the static cases where all of the data 
is available at training time, as well as in the dynamic case where the data 
is added over time. These approximations can be divided into two distinct 
tracks, inducing inputs and basis function expansion. A comprehensive review of 
inducing input approaches can be found in 
\cite{Quinonero-Candela2005:SparseGPregression}. An introduction to basis 
function expansion can be found in \cite{RasmussenWilliams2006:GPforML, 
Svensson2016, Solin2019, Kok2018}. \par
Here, each \abbrGP will be approximated using inducing inputs, which is a set 
of discrete points fixed at some arbitrary location. Here, the points are 
fixed at $\indi_l^\indu$ for ${l=1,\dots,L}$ and their corresponding function 
values are given by $\indub_l=\begin{bmatrix}\indu^1(\indi_l^\indu) & 
\indu^2(\indi_l^\indu) & \cdots & \indu^J(\indi_l^\indu)\end{bmatrix}^T$ and are
referred to as inducing inputs. Gather the inducing inputs 
in the vector ${\Xi=\begin{bmatrix}\indub_1^T & 
\indub_2^T & \cdots & \indub_L^T
\end{bmatrix}^T}$ and their corresponding coordinates\- in 
$\indil=\begin{bmatrix}\indi_1^\indu & \indi_2^\indu & \cdots & 
\indi_L^\indu
\end{bmatrix}^T$. Next, let 
\begin{align}
\Kuu &= \K[](\indil, \indil) = \begin{bmatrix}
k(\indi_1^\indu, \indi_1^\indu) & \cdots & 
k(\indi_1^\indu,\indi_L^\indu)\\
\vdots & \ddots & \vdots\\
k(\indi_L^\indu, \indi_1^\indu) & \cdots & 
k(\indi_L^\indu,\indi_L^\indu)
\end{bmatrix},\\
\K[\bdot\indu]^k &= \K[](\indi_k, \indil) = 
\begin{bmatrix}
k(\indi_k, \indi_1^\indu) & \cdots & k(\indi_k, \indi_L^\indu)
\end{bmatrix},\\
\Ktilde[] &= \K[]\otimes I,
\end{align}
where $I$ is the identity matrix. If the inputs are assumed independent of each 
other and the same kernel is used for each input $\indu^j$, the complete 
covariance matrix for $\indul$ is given by $\Ktilde$. By using the Fully 
Independent Conditional (\abbrFIC) approximation 
\cite{Quinonero-Candela2005:SparseGPregression, Snelson2005:SPGP}, the model 
can now be written as
\begin{subequations}
	\begin{align}
	\state_k &= F_k \state_{k-1} + G_k 
	(\Ktilde[\bdot\indu]^{k} 
	\Kuutilde^{-1}\indul+\vec{v}_k)\\
	\obs_k &= H_k \state_k + \vec{e}_k\\
	\indul &\sim \Ndist(0, \Kuutilde)
	\end{align}
\end{subequations}
where $\vec{v}_k$ is Gaussian white noise with covariance matrix ${\Lambda_k = 
k(\indi_{k},\indi_{k})- \K[\bdot\indu]^k 
\Kuu^{-1}[\K[\bdot\indu]^k]^T}$.
Hence, the input at a given point $\vec{z}_k$ is given by a linear combination 
of the inducing inputs. Lastly, the matrix $\Kuutilde$ is not always well 
conditioned so to avoid numerical issues define 
\begin{equation}
\iindul = \Kuutilde^{-1}\indul = \begin{bmatrix}
\iindub_1^T & \iindub_2^T & \cdots & \iindub_L^T
\end{bmatrix}^T
\end{equation}
and note that 
\begin{equation*}
\mathrm{Cov}\left[\iindul\right]=\mathrm{Cov}\left[\Kuutilde^{-1}\indul\right]=\Kuutilde^{-1}\Kuutilde\Kuutilde^{-T}=\Kuutilde^{-1},
\end{equation*}
which is only needed for interpreting the covariance of the estimates and is 
not necessary for implementing the estimation. To allow the model to adapt to 
changes in the input, the inducing point state will be modeled as a random walk 
process. That is,
\begin{equation}
	\iindul_k = \iindul_{k-1}+\acute{\vec{w}}_k
\end{equation}
where $\acute{\vec{w}}_k$ is Gaussian white noise with covariance matrix 
$\Sigma$. The state vector is then augmented with $\iindul$ and the model 
becomes
\begin{subequations}
	\label{eq:approximateGPmodel}
	\begin{align}
	\state_k &= F_k \state_{k-1} + G_k 
	(\Ktilde[\bdot\indu]^{k} 
	\iindul_{k-1}+\vec{v}_k)\\
	\iindul_k &= \iindul_{k-1}+\acute{\vec{w}}_k\\
	\obs_k &= H_k \state_k + \vec{e}_k,
	\end{align}
\end{subequations}
and note that $\Ktilde[\bdot\indu]^k$ still depends on the parameters $\vec{z}$.

\subsection{State-dependent input}
In the considered scenario, the vehicle acceleration depends on the location of 
the vehicle within the intersection. Hence, the input depends on the position 
of the vehicle, i.e., 
\begin{subequations}
	\begin{align}
	\label{eq:statedependentinput}
		\indi_k&=D_k\state_{k-1}\\
		D_k &= \begin{bmatrix}
		I & 0
		\end{bmatrix}.
	\end{align}
\end{subequations}
The acceleration of course also depends on the velocity of the 
vehicle (The centripetal acceleration through a curve is given by 
$a_c=v^2/R$, where $R$ is the radius of the curve and $v$ is the 
absolute speed.). However, as this would quadratically scale the inducing point 
space $\indi^\indu_l$ (from $\R^2$ to $\R^4$), this is neglected for 
computational 
reasons. The full model is then described by
\begin{subequations}
	\label{eq:fullmodel}
	\begin{align}
	\state_k &= F_k \state_{k-1} + G_k\label{eq:model:statedynamics} 
	(\Ktilde[\bdot\indu]^{k} 
	\iindul_{k-1}+\vec{v}_k)\\\label{eq:model:inputdynamics}
	\iindul_k &= \iindul_{k-1} + \acute{\vec{w}}_k\\
	\indi_k &= D_k\state_{k-1}\\
	\obs_k &= H_k \state_k + \vec{e}_k.
	\end{align}
\end{subequations}
Recall that $\Ktilde[\bdot\indu]^{k}=\K[](\vec{z}_k, 
\indil)\otimes I$ and note that the model is now nonlinear in 
the states due to the dependence of $\K[](\vec{z}_k, 
\indil)$ on $\state_{k-1}$.

\subsection{Kernel choice}
The choice of kernel function specifies what family of functions the \abbrGP is 
able to approximate well \cite{RasmussenWilliams2006:GPforML}. A common choice 
is the squared exponential kernel
\begin{equation}
\label{eq:squaredexponentialkernel}
k(\vec{z},\vec{z}^*) = 
\sigma_f^2\exp\left(-\frac{1}{2l^2}||\vec{z}-\vec{z}^*||^2\right)
\end{equation}
which will be used here as well. The hyperparameters $\theta=(\sigma_f^2,~l)$ 
govern the properties of the kernel, where $\sigma_f^2$ controls the general 
variance and $l$ is the characteristic length-scale and controls the 
width of the kernel. The hyperparameters can either be learned online 
\cite{Huber2014:RGP}, \cite{Titsias2009:VIofInducingVariables} or selected 
manually based on insight about the physical properties of the input.

\section{Estimation}\label{sec:estimation}

The model \eqref{eq:fullmodel} is nonlinear in the states and can be 
recursively estimated using, e.g., an 
\abbrEKF based on a first order Taylor expansion 
\cite{Sarkka2010:BayesianFiltering}. The \abbrEKF assumes that the prediction 
and filter distributions are both normally distributed as 
\begin{align}
\begin{pmatrix}
	\state_{k+1|k}\\
	\iindul_{k+1|k}
\end{pmatrix}&\sim\Ndist\left(\begin{pmatrix}
	\hat{\state}_{k+1|k}	\\
	\hat{\iindul}_{k+1|k}
\end{pmatrix}, \begin{pmatrix}
	\vec{P}^x_{k+1|k} & \vec{P}^{x\iindu}_{k+1|k}\\
	\vec{P}^{\iindu x}_{k+1|k} & \vec{P}^\iindu_{k+1|k}
\end{pmatrix}\right)\\
\begin{pmatrix}
\state_{k|k}\\
\iindul_{k|k}
\end{pmatrix}&\sim\Ndist\left(\begin{pmatrix}
\hat{\state}_{k|k}	\\
\hat{\iindul}_{k|k}
\end{pmatrix}, \begin{pmatrix}
\vec{P}^x_{k|k} & \vec{P}^{x\iindu}_{k|k}\\
\vec{P}^{\iindu x}_{k|k} & \vec{P}^\iindu_{k|k}
\end{pmatrix}\right).
\end{align}
By using the upper triangular structure of the state transition model 
\eqref{eq:model:statedynamics} and \eqref{eq:model:inputdynamics}, the \abbrEKF 
time-update becomes \cite{Veiback2019:EKFGP}
\begin{subequations}
	\begin{align}
		\hat{\state}_{k+1|k} &= 
		F_k\hat{\state}_{k|k}+G_k\Ktilde[\bdot\indu]
		(D_k\hat{\state}_{k|k})\hat{\iindul}_{k|k}\\
		\hat{\iindul}_{k+1|k} &= \hat{\iindul}_{k|k}\\
		P_{k+1|k}^x &= 
		\vec{F}_xP_{k|k}^x\vec{F}_x^T+\vec{F}_xP^{x\iindu}_{k|k}\vec{F}_\iindu^T+
		\vec{F}_\iindu P^{\iindu x}_{k|k}\vec{F}_x^T\nonumber\\
		&\quad + \vec{F}_\iindu 
		P^\iindu_{k|k}\vec{F}_\iindu^T+G_k\Lambda_{k}G_k^T\\
		P_{k+1|k}^{x\iindu} &= \vec{F}_xP^{x\iindu}_{k|k}+\vec{F}_\iindu 
		P^\iindu_{k|k}\\
		P_{k+1|k}^{\iindu x} &= (P^{x\iindu}_{k+1|k})^T\\
		P_{k+1|k}^\iindu &= P_{k|k}^\iindu+\Sigma
	\end{align}
\end{subequations}
and the measurement update becomes
\begin{subequations}
	\begin{align}
		S_k &\defas R+H_kP^x_{k|k-1}H_k^T\\
		L_k^x &\defas P^x_{k|k-1}H_k^TS_k^{-1}\\
		L_k^\iindu &\defas P^{\iindu x}_{k|k-1}H_k^TS_k^{-1}\\
		\hat{\state}_{k|k} &= 
		\hat{\state}_{k|k-1}+L_k^x(\obs_k-H_k\hat{\state}_{k|k-1})\\
		\hat{\iindul}_{k|k} &= 
		\hat{\iindul}_{k|k-1}+L_k^\iindu(\obs_k-H_k\hat{\state}_{k|k-1})\\
		P^x_{k|k} &= P^x_{k|k-1} - L^x_kS_k(L^x_k)^T\\
		P^{x\iindu}_{k|k} &= P^{x\iindu}_{k|k-1} - L^x_kS_k(L^\iindu_k)^T\\
		P^{\iindu x}_{k|k} &= (P^{x\iindu}_{k|k})^T\\
		P^\iindu_{k|k} &= P^\iindu_{k|k-1} -L^\iindu_kS_k(L^\iindu_k)^T,
	\end{align}
\end{subequations}
where
\begin{align}
\vec{F}_x &= 
F_k-\frac{G_k}{l^2} \left(\sum_{l=1}^{L}
k(\vec{z}_k,\indi_l^\indu)\cdot(\vec{z}_k-\indi_l^\indu)^T\iindub_l\right)D_k\\
\vec{F}_\iindu &= G_k\Ktilde[\bdot\indu]^{k}.
\end{align}
See Appendix \ref{app:sec:jacobian} for a derivation.

\section{Simulation and Results}\label{sec:simulationandresults}

	To illustrate the application of the proposed estimation approach and the 
	\abbrGPASSM, position observations from a set of vehicles passing through 
	the intersection illustrated in Fig \ref{fig:description:map} were 
	simulated. 
	\subsection{Simulation Parameters}
	There are a number of parameters to be either learned or chosen. Here, they 
	are all manually selected based on prior knowledge of the physical 
	properties of the scenario. For the vehicle motion, a \abbrCV is chosen, 
	i.e., 
	\begin{equation}
	\label{eq:model:dynamicmodel}
	F_k = \begin{bmatrix}
	1 & T\\
	0 & 1
	\end{bmatrix}\otimes I \qquad
	G_k = \begin{bmatrix}
	T^2/2\\
	T
	\end{bmatrix}\otimes I,
	\end{equation}
	where $T$ is the sampling interval. For the observation model, it is 
	assumed that the position of the vehicle is measurable with some noise, 
	i.e., 
	\begin{equation}
	H_k = \begin{bmatrix}
	I & 0
	\end{bmatrix}\qquad R=\sigma_e^2 I,
	\end{equation}
	where $\sigma_e^2$ is the measurement noise variance. \par
	As for the \abbrGP, there are three parameters to be chosen: the location 
	of the inducing points $\indi_l^\indu$, the kernel variance $\sigma^2_f$, 
	and the kernel length scale $l$. The inducing points are fixed on a grid 
	covering the entire intersection, see Fig~\ref{fig:description:map}, 
	uniformly spaced using a grid spacing 
	$\delta_{\indi_l^\indu}=\SI{1}{\meter}$, which was chosen as a trade-off 
	between accuracy and computational burden. The kernel variance and length 
	scale are chosen under the notion that the acceleration of a vehicle is 
	a local phenomena and varies quickly over distance/time. The length scale 
	is thus chosen as $l=0.5$ and the variance as $\sigma^2_f=0.05$. The 
	simulation parameters are summarized in Table 
	\ref{tab:three-way-intersection:simulationparameters}.\par
	During the simulations, all the vehicles were initiated to the true initial 
	position and velocity. In total $M=100$ simulations were run. 
	\begin{table}[t]
		\centering
		\caption{Simulation parameters for three-way intersection scenario}
		\label{tab:three-way-intersection:simulationparameters}
		\begin{tabular}{c|c|c}
			\textbf{Parameter} & \textbf{Description} & 
			\textbf{Value}\\\hline&&\\[-1em]
			L & \# Inducing points & 310\\
			N & \# Vehicles & 30 \\
			$f_s$ & Sampling rate & $\SI{2}{\hertz}$\\
			$\sigma_e^2$ & Measurement noise variance & 0.2\\
			$\sigma^2_f$ & Kernel variance & 0.05\\
			$l$ & Kernel length scale & 0.5\\
			R & \abbrEKF measurement noise variance & 1\\
			$\delta_{\indi_l^\indu}$ & Grid spacing & $\SI{1}{\meter}$
		\end{tabular}
	\end{table}
	\begin{figure}[h]
		\centering
		\includegraphics[width=\columnwidth]{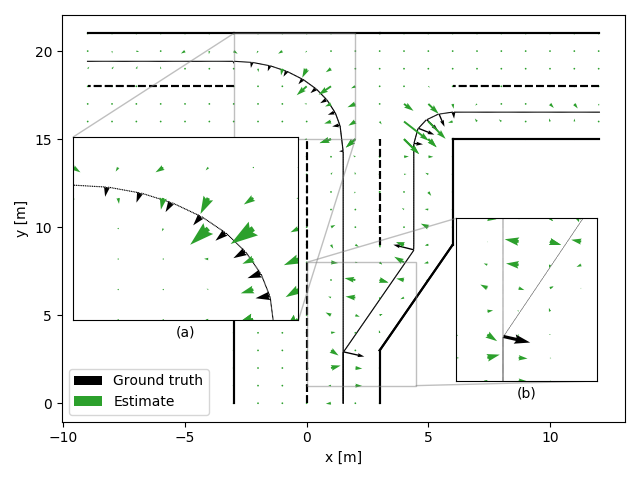}
		\caption{Acceleration estimated by the \abbrGPASSM for one simulation 
			with ground truth for comparison. Ground truth is plotted over the 
			paths, estimates are over the sparse grid approximation. The 
			estimated accelerations mimic the true accelerations well, 
			see zoomed in area (a). At the path split point, see zoomed in area 
			(b), the estimated accelerations diverge from the true.}
		\label{fig:three-way-intersection:accelerationestimate}				
	\end{figure}
	\subsection{Qualitative Analysis}
\begin{figure}[h]
		\centering
		\includegraphics[width=\columnwidth]{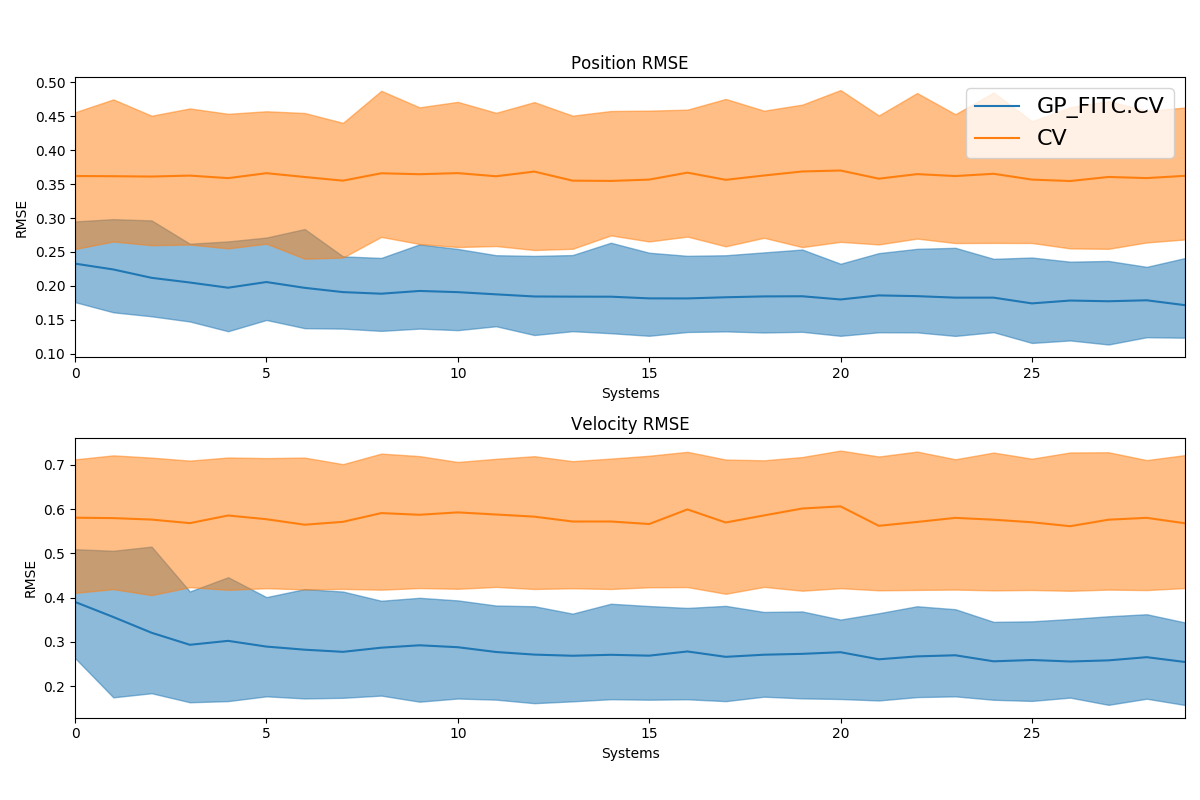}
		\caption{\abbrRMSE over vehicles. Later vehicles use the accelerations 
			learned from previous vehicles and thus give a lower \abbrRMSE for 
			the \abbrGPASSM, but not for the \abbrCV model. Confidence bands 
			are given by the $2.5$th and $97.5$th percentile over the 
			simulations.}%
		\label{fig:three-way-intersection:rmse}%
		\vspace*{\floatsep}%
		\includegraphics[width=\columnwidth]{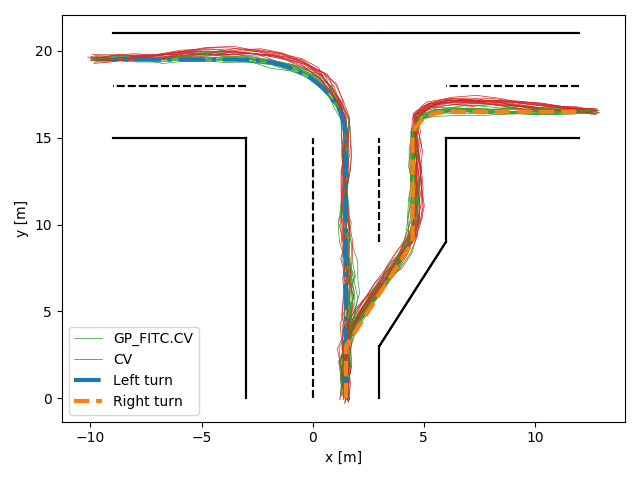}
		\caption{50 vehicle trajectories of the two models. The \abbrGPASSM is 
			learned from a training data set of 30 vehicles. The \abbrCV model 
			suffers from bias, which the \abbrGPASSM alleviates. That is, the 
			green trajectories accurately follow the paths while the red do 
			not.}%
		\label{fig:three-way-intersection:drytrajectories}%
	\end{figure}

\begin{figure*}[b]
		\centering
		\subfloat[\label{fig:three-way-intersection:errors:iteration_0}The 
		first vehicle on each path]{\includegraphics[width=\textwidth, 
			height=.45\textheight]{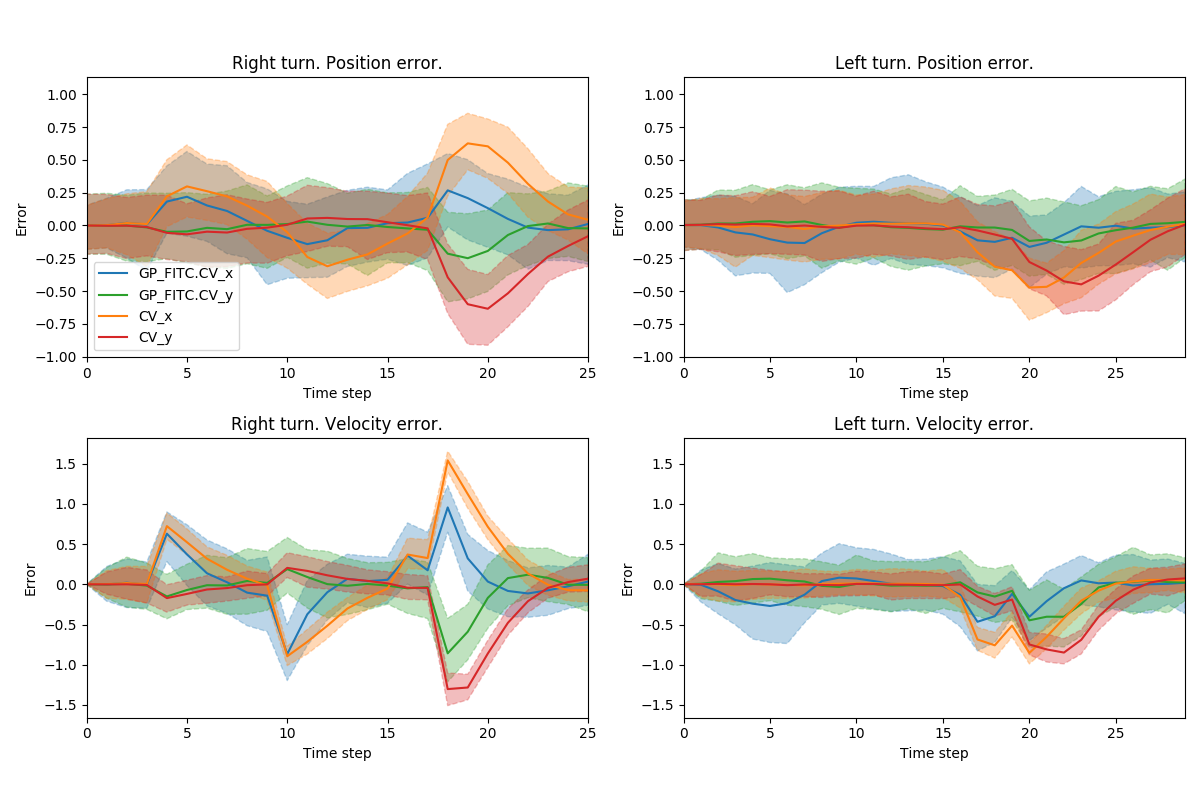}}\\%
		\subfloat[\label{fig:three-way-intersection:errors:iteration_-1}The 
		last vehicle on each path]{\includegraphics[width=\textwidth, 
			height=.45\textheight]{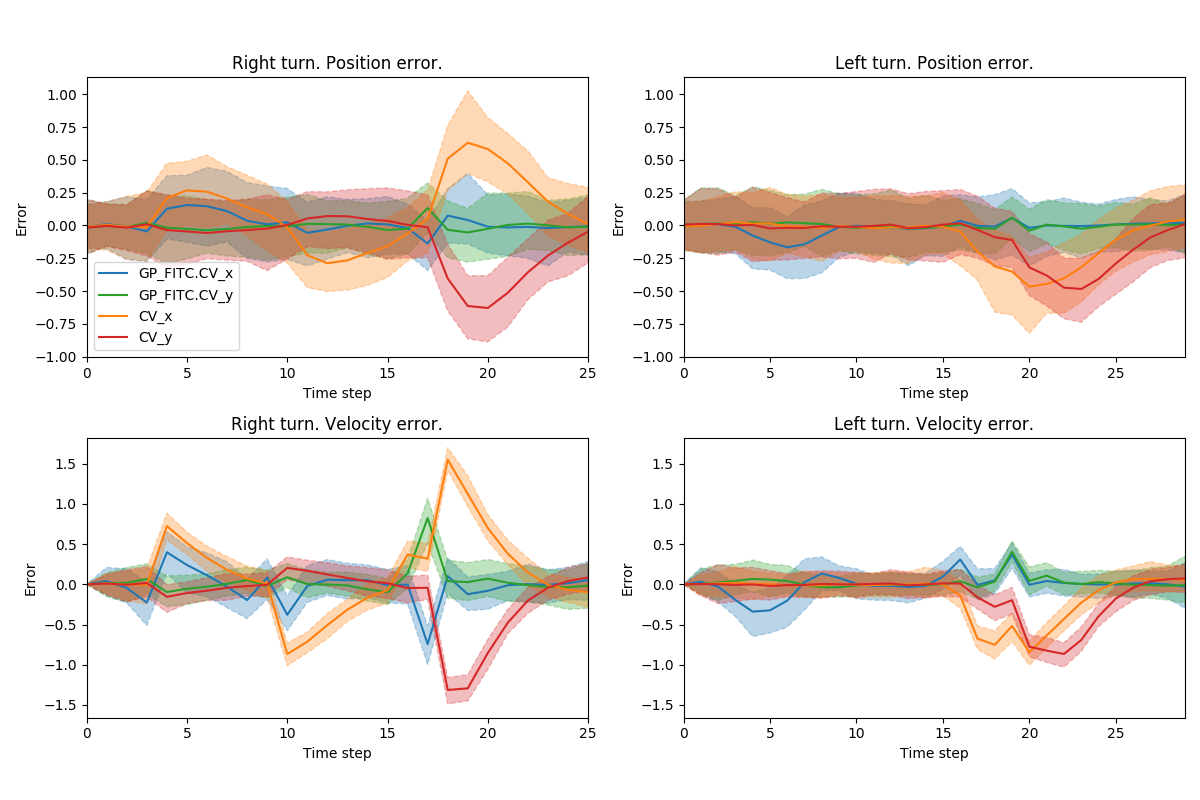}}%
		\caption{State errors for two different vehicles for both the 
			\abbrGPASSM and \abbrCV model. 	Errors are separated by path and by 
			dimension. Confidence bands are	given by the $2.5$th and $97.5$th 
			percentile over the simulations. The two subfigures depict the 
			first 
			and the last vehicle on each path, i.e., it is not necessarily the 
			first and last vehicle in total.}
				\label{fig:errors}
	\end{figure*}

	As a baseline, a \abbrCV model without the input learning was used. The 
	Root Mean Squared Error (\abbrRMSE) was calculated for each vehicle in each 
	separate simulation and is visualized in Fig 
	\ref{fig:three-way-intersection:rmse}. 
	Fig~\ref{fig:three-way-intersection:drytrajectories} visualizes 
	$50$ trajectories of a \abbrCV, as well as of the \abbrGPASSM where the 
	input has been learned in advance using $30$ vehicles. The state 
	errors for the first vehicle on each path are given in 		
	Fig~\ref{fig:errors}\subref{fig:three-way-intersection:errors:iteration_0}
	 and the last on 
	each path in 
	Fig~\ref{fig:errors}\subref{fig:three-way-intersection:errors:iteration_-1}.
	Note that this is not necessarily the first or last vehicle in total and 
	also that there is no guarantee that the same number of vehicles have 
	traversed each path.
	\par
	From 
	Fig~\ref{fig:errors}\subref{fig:three-way-intersection:errors:iteration_0}
	it is evident that already for the first vehicle, there are benefits of 
	including the \abbrGP. Before any accelerations are experienced, the model 
	mimics the \abbrCV model exactly, but as accelerations come into play 
	(around time step 3--4) the proposed model improves over the standard 
	\abbrCV. This is due to the process noise being inherently nonwhite, which 
	the \abbrGP captures. As the number of vehicles increases, see 
	Fig~\ref{fig:errors}\subref{fig:three-way-intersection:errors:iteration_-1},
	 the \abbrGPASSM 
	learns the accelerations required to stay on the trajectory; see time step 
	15--25 for both paths. Lastly, even larger discrepancies between the 
	\abbrGPASSM and \abbrCV model would be evident if $k$-step ahead prediction 
	was used, since the \abbrCV would continue in a straight path and the 
	\abbrGPASSM would follow the average trajectory of the past vehicles.\par
	Now, there are some peculiarities. For instance, see 
	Fig~\ref{fig:errors}\subref{fig:three-way-intersection:errors:iteration_-1}
	where the \abbrGPASSM is actually worse than the \abbrCV model 
	between time steps 3--10. This is caused by the 
	acceleration discontinuity where the two paths split. This is also evident 
	in Fig~\ref{fig:three-way-intersection:accelerationestimate}, (zoomed in 
	area (b)), where the discontinuity causes a lot of small sideways 
	accelerations where the \abbrGP is compensating for its own errors in a 
	sense. Fig~\ref{fig:three-way-intersection:accelerationestimate} also 
	indicates that the learned acceleration mimic the true closely, see zoomed 
	in area (a).\par
	From Fig~\ref{fig:three-way-intersection:drytrajectories} it is evident 
	that the \abbrGPASSM follows the two paths better than the \abbrCV 
	model. Whereas the \abbrCV model has a clear bias during the turns, the 
	\abbrGPASSM does not suffer from this.

\section{Conclusion and Future Work}\label{sec:conclusionandfuturework}

	A Gaussian Process Augmented State-Space Model has been proposed for 
	learning unknown, but common, accelerations of vehicles through static 
	environments. The model generalizes to cases where a simple motion model is 
	sought after, but where the bias associated with such are not. The model 
	was shown to improve over an ordinary Constant Velocity (\abbrCV) model and 
	removed the bias when the accelerations were non-zero. An issue with the 
	model is that it can not handle ambiguities in the input it is trying to 
	learn and will in some of these cases perform worse than a \abbrCV model. 
	The model is, however, attractive as it allows a simple motion model to be 
	used in combination with a data-driven model for learning unknown 
	characteristics of the underlying system online. It also facilitates 
	distributed learning of unknown characteristics between systems that behave 
	similarly. The learned input function itself 
	might also be of use since its physical interpretation in many cases is 
	easily found.\par
	For the model to reach its full potential, the input ambiguities must be 
	addressed. It is also necessary to find an efficient way to factorize the 
	input space so as 
	to reduce the computational burden, e.g., through dividing the area of 
	interest into hexagonal surfaces \cite{Kok2018}. Moreover, the 
	approximation strategy of the Gaussian Process needs to be evaluated. If an 
	inducing input approach is used, methods to add, remove, or move these 
	online is necessary for reducing computational burden and to enable the 
	model to be used in large-scale applications. 

\newpage
\section{Acknowledgments}\label{sec:acknowledgment}

This work was partially supported by the Wallenberg AI,
Autonomous Systems and Software Program (\textsc{WASP}) funded
by the Knut and Alice Wallenberg Foundation.

\appendices
\section{Derivation of model Jacobian}\label{app:sec:jacobian}
\renewcommand{\theequation}{\thesection.\arabic{equation}}
\setcounter{equation}{0}

Given the model
\begin{subequations}		
	\label{app:eq:fullmodel}
	\begin{align}
	\state_k &= F_k \state_{k-1} + G_k 
	(\Ktilde[\bdot\indu]^{k} 
	\iindul_{k-1}+\vec{v}_k)\\
	\iindul_k &= \iindul_{k-1} + \vec{w}_k\\
	\vec{z}_k &= D\state_{k-1}
	\end{align}
\end{subequations}
the partial derivative of $\state_k$ with respect to $\state_{k-1}$ is given by
\begin{align}
\label{app:eq:dxdx}
\frac{\partial \state_k}{\partial \state_{k-1}} &= F_k+G_k 
\left(\frac{\partial \Ktilde[\bdot\indu]^{k}}{\partial 
	\state_{k-1}}\iindul_{k-1} + \frac{\partial \vec{v}_k}{\partial 
	\state_{k-1}}\right)
\end{align}
where the derivative of the squared exponential kernel is given by
\begin{equation}
\frac{\partial k(\vec{z},\vec{z}^*)}{\partial \vec{z}} = 
-\frac{k(\vec{z},\vec{z}^*)}{l^2}(\vec{z}-\vec{z}^*)^T
\end{equation}
and the derivative of the noise component $\vec{v}_k$ is given by
\begin{equation}
\frac{\partial \vec{v}_k}{\partial \state_{k-1}} = \frac{\partial 
	\vec{v}(D_k\state_{k-1})}{\partial \state_{k-1}} = 
\frac{1}{2}\frac{\vec{v}_k}{\Lambda_k}\frac{\partial \Lambda_k}{\partial 
	\state_{k-1}}D_k.
\end{equation}
For proof, see Appendix B in \cite{Veiback2019:EKFGP}. Now, \eqref{app:eq:dxdx} 
can be written
\begin{align}
\vec{F}_x \defas \frac{\partial \state_k}{\partial \state_{k-1}} &= 
F_k+G_k\bigg(
-\frac{1}{l^2}\sum_{l=1}^{L}[
k(\vec{z}_k,\indi_l^\indu)(\vec{z}_k-\indi_l^\indu)^T\vec{w}_l]\nonumber\\
&\quad+ 
\frac{1}{2}\frac{\vec{v}_k}{\Lambda_k}\frac{\partial\Lambda_k}{\partial\state_{k-1}}\bigg)D_k
\label{app:eq:fulldxdx}
\end{align}
Furthermore the derivative of $\state_k$ w.r.t. $\iindul_{k-1}$ is given by
\begin{equation}
\vec{F}_\iindu \defas \frac{\partial\state_k}{\partial\iindul_{k-1}} = 
G_k\Ktilde[\bdot\indu]^{k}.
\end{equation}

\bibliographystyle{IEEEtran}
\bibliography{library}
\end{document}